\documentclass{ws-ijmpa}

\begin{document}

\markboth{Bing An Li}
{Instructions for Typing Manuscripts (Paper's Title)}

%
\catchline{}{}{}{}{}
%

\title{CURRENT ALGEBRA BASED EFFECTIVE CHIRAL THEORY OF MESONS AND A NEW EW THEORY
}

\author{Bing An Li}

\address{Department of Physics, Univ. of Kentucky, Lexington,
KY, 40506, USA}

\maketitle

\pub{Received (Day Month Year)}{Revised (Day Month Year)}

\begin{abstract}
A current algebra based effective chiral theory of pseudoscalar, vector, axial-vector
mesons is reviewed. A new mechanism generating the masses and guage fixing terms 
of gauge boson is revealed from this effective theory. A EW theory without Higgs
is proposed. The masses and gauge fixing terms of W and Z are dynamically generated.
Three heavy scalar fields are dynamically generated too. They are ghosts. 

\end{abstract}
Current algebra is a very successful theory of pseudoscalar, vector, and axial-vector mesons
at low energies. Based on current algebra the chiral theory of pseudoscalar,
vector, and axial-vector mesons is constructed[1]
\[{\cal L}=\bar{\psi}(x)(i\gamma\cdot\partial+\gamma\cdot v
+\gamma\cdot a\gamma_{5}
-mu(x))\psi(x)
-\bar{\psi(x)}M\psi(x)\]
\[+{1\over 2}m^{2}_{0}(\rho^{\mu}_{i}\rho_{\mu i}
+\omega^{\mu}\omega_{\mu}+a^{\mu}_{i}a_{\mu i}+f^{\mu}f_{\mu})\]
where \(a_{\mu}=\tau_{i}a^{i}_{\mu}+f_{\mu},\)
\(v_{\mu}=\tau_{i} 
\rho^{i}_{\mu}+\omega_{\mu},\)
\(u=exp\{i\gamma_{5}(\tau_{i}\pi_{i}+
\eta)\}.\) The theory has been extended to three flavors.
In this theory vector and axial-vector mesons are defined
by corresponding quark vector and axial-vector currents, pseudoscalar
mesons are introduced by nonlinear $\sigma$ model. Therefore, this theory
is the filed theory realization of current algebra.
Integrating out quark fields the effective Lagrangian of mesons is derived.
As a matter of fact, the meson Lagrangian is the effect of one quark loop
diagrams(in Euclidean space).

This effective theory has
dynamical chiral symmetry breaking
\(<\bar{\psi}\psi>\sim m^3,\)
and
explicit chiral symmetry in the limit \( m_q\rightarrow 0\).
A cut-off \(\Lambda\sim1.8GeV\) is introduced.
In chiral limit there are two parameters: $f_\pi$ and a universal
coupling constant g which is determined to be 0.39 by fitting $\rho\rightarrow ee^+$. 
They are defined by loop integrals. Both $f^2_\pi$ and $g^2$
are proportional to $N_C$. The order of physical meson fields in $N_C$ expansion
are defined to be $O(\sqrt{N_C})$. Therefore $N_C$ expansion is carried out in this 
theory. Tree diagrams are at the leading order and loop diagrams of mesons are at higher order.
The mass difference of $\omega-\rho$ and $a_1-f_1$ and $\phi\rightarrow\rho\pi$ are at next
leading order of $N_C$ expansion. $N_C$ expansion governs the calculations of this theory.
 
Because of the cancellation of the two diagrams obtained from
\[-im\bar{\psi}\gamma_5\tau^i\psi\pi^i\;\;\;{1\over2}m\bar{\psi}\psi\pi^2\]
the Goldstone theorem is revealed,
\[m^{2}_{\pi}=-{2\over f^{2}_{\pi}}(m_{u}+m_{d})<0|\bar{\psi}\psi|0>.\]
The KSFR sum rule leads to
\[m^{2}_{\rho}=2{f^{2}_{\pi}\over g^{2}}=6m^2.\] 
A modified Weinberg's second sum
rule is obtained
\[(1-{1\over 2\pi^2g^2})m^2_a=2m^2_{\rho}.\]
The pion mass is from the explicit chiral symmetry breaking. The $m_\rho$ originates in 
dynamical chiral symmetry breaking. The $m_{a_1}$ is obtained from a new mechanism
of chiral symmetry breaking. This new mechanism is a new discovery from this current algebra
based effective chiral field theory.

The strong decay widths are calculated
\(\Gamma_\rho=150MeV,\;\;\Gamma_{K^*}=45MeV,\;\;\Gamma_\phi=4.2MeV,\;\;
\Gamma_a=326MeV...\). EM and weak decays are calculated too. Theory agrees with data.
Weinberg's first sum rule is satisfied analytically. In the amplitude of 
$a_1\rightarrow\rho\pi$ there are two terms. The theoretical ratio of d-wave to s-wave 
agrees with data well. There is strong cancellation between the two terms. In the limit
of soft pion the relation found by current algebra is satisfied. The puzzle of current algebra
for long time is solved. The decay widths of $K_1(1400)$ and $f_1(1420)$ are much 
narrower than $a_1$. The theory provides the explanation and numerical results agree
with data. 

The scattering lengths and slopes $\pi-\pi$ scattering are derived and they are the same as
obtained by Weinberg. The amplitudes of \(I=1,2\) agree with data. The s-wave(\(I=1\)) agrees
with data at lower energies and the contribution of $\sigma$ meson must be taken into account.
$\pi-K$ scattering has been studied too. Theory agrees with data very well.
Besides the $\rho$ pole an intrinsic form form factor is found in $\pi$ form factor. This new
form factor redeems the shortcomings of the $\rho$ pole form factor.
\[f_\pi(q^2)=\frac{1+\frac{q^2}{2\pi^{2}f^{2}_{\pi}}
[(1-{2c\over g})^{2}-4\pi^{2}c^{2}]}{q^2-m^2_\rho+i\sqrt{q^2}\Gamma_\rho(q^2)}\]
where \(c=\frac{f^2_\pi}{2gm^2_\rho}\). This form factor agrees with data in both
time-like and space-like regions. 

The kaon form factors are studied too. Theory agrees with data very well.

PCAC is satisfied. Many $\tau\rightarrow mesons+\nu$ are calculated. Theory agrees
with data.

The Lagrangians of normal parity and Wess-Zumino-Witten anomaly are derived from
the same Lagrangian. The two parameters of WZW anomaly are determined.
The anomalous L related to $\omega$ is derived as
\begin{eqnarray}
{\cal L}=
\frac{N_{c}}{(4\pi)^{2}}{2\over 3}\varepsilon^{\mu\nu\alpha\beta}
\omega_{\mu}Tr\partial_{\nu}UU^{\dag}\partial_{\alpha}UU^{\dag}
\partial_{\beta}UU^{\dag}
+\frac{2N_{c}}{(4\pi)^{2}}\varepsilon^{\mu\nu\alpha\beta}
\partial_{\mu}\omega_{\nu}Tr\{i[\partial_{\beta}UU^{\dag}\nonumber \\
(\rho_{\alpha}+a_{\alpha})-\partial_{\beta}U^{\dag}U(\rho_{\alpha}
-a_{\alpha})]
-2(\rho_{\alpha}+a_{\alpha})U(\rho_{\beta}-a_{\beta})
U^{\dag}-2\rho_{\alpha}a_{\beta}\}.\nonumber
\end{eqnarray}

The low energy theorem of $\gamma\rightarrow3\pi$ which is less than data by $30\%$
is corrected[2]
\[A_{\gamma3\pi}(0,0,0)=\frac{2e}{\pi^2 f^3_\pi}(1+\frac{6c^2}{g^2})
=12.2GeV^3,\]
it agrees with data. The cross section of $\pi+(A,Z)\rightarrow\pi\pi+(A,Z)$ is calculated
\(\sigma=1.34nb\).
It agrees with data.

$ee^+\rightarrow\pi^0\pi^+\pi^-$ is dominant by the anomaly $\gamma\pi\pi\pi$.
The contribution of the WZW anomaly(leading order) is far too small. The next
leading order has to be taken into account. This effective chiral theory can go
beyond the WZW anomaly and the next leading terms are found. 
Theory agrees with data
very well. There are many other anomalous processes in which next leading terms
must be taken into account.

{\bf Comparison with other chiral theories of mesons}\\
1)The Chiral Perturbation Theory is the low energy limit of this effective
theory, all the 10 coefficients are predicted
2)Hidden gauge theory. All the 13 parameters of the hidden gauge theory are predicted.
Three new terms are found[3].
3)Nambu-Jano-Lasinio model is a model of four quark interactions. This effective theory
is not. There are highly nonlinear quark interactions.
4)In 1966[4] we
 first used the
effective L of four quark interactions to study the physics of mesons and baryons.
We have obtained a mass relations between meson and baryons
\[\frac{m^2_K-m^2_\pi}{m^2_\rho-m^2_\pi}={3\over4}\frac{M_{\Sigma^*}-M_\Delta}{M_\Delta
-M_N}\]
Based on the studies in this paper the strong interactions
\(e_V\bar{\psi}Y\gamma_\mu\psi B_\mu\),
\(e_A\bar{\psi}Y\gamma_\mu\gamma_5\psi C_\mu\).
are proposed.

{\bf New symmetry breaking and a possible new EW theory}\\
In chiral limit, the original L has global chiral symmetry, $m_a=m_\rho$.
After the quark loops are taken into account
the mass formula of $a_1$
is obtained, which agrees with data.
The global chiral symmetry is broken. Additional mass term and
\((\partial_\mu a_\mu)^2(gauge\;fixing\;term)\) are dynamically generated
by the vacuum polarization of $a_1$-fields. The gauge fixing term is related to the
new factor in $a_1$ mass formula.
The same happens to the vacuum polarization of
charged vector currents.
Therefore, a new mechanism upon which
masses of gauge bosons and the gauge fixing terms can be dynamically generated
is revealed from this effective field theory.
It is known that scalar fields are required to cancel $\frac{k_\mu k_\nu}{m^2}$
after the intermediate bosons gain masses. The Higgs mechanism satisfies these
two requirements. However, they are added by hand and they cause other problems.
So far, Higgs has not been found.
In Res.[5] the new mechanism found in the effective field theory
has been used to construct a new Electro-Weak theory without Higgs.
One neutral and charged scalar fields are dynamically generated.
EW theory without Higgs can be expressed as
\[{\cal L}=
-{1\over4}A^{i}_{\mu\nu}A^{i\mu\nu}-{1\over4}B_{\mu\nu}B^{\mu\nu}
+\bar{q}_L\{i\gamma\cdot\partial
+{g\over2}\tau_{i}\nonumber \\
\gamma\cdot A^{i}+g'{Y\over2}\gamma\cdot B\}
q_{L}
+\bar{q}_{R}\{i\gamma\cdot\partial+g'{Y\over2}\gamma\cdot B\}q_{R}\]
\[-m_q\bar{q}q\]
There are neutral axial-vector and charged vector and axial vector currents of fermions.
One-loop vacuum polarization of Z-fields is expressed as
\[\Pi^Z_{\mu\nu}={1\over2}F_{Z1}(z)(p_\mu p_\nu-p^2 g_{\mu\nu})+F_{Z2}(z)
p_\mu p_\nu+{1\over2}\Delta m^2_Z g_{\mu\nu},\]
where $F_2$ is finite and $\Delta m^2_Z$ is a constant. In the limit of zero fermion mass,
these two terms disappear[5]. The vacuum polarization of W-fields has similar expression.
The masses of W and Z and gauge fixing terms are dynamically generated by corresponding
vacuum polarization diagrams of W and Z. The gauge fixing terms lead to dynamical
generation of scalar fields
\[Z_\mu=Z'_\mu+{1\over m_{\phi_Z}}\partial_\mu\phi_Z,\;\;\partial_\mu Z'_\mu=0,\;\;
W_\mu=W'_\mu+{1\over m_{\phi_W}}\partial_\mu\phi_W,\;\;\partial_\mu W'_\mu=0.\] 
Their masses are determined to be
\(m_{\phi_Z}=3.78\times10^{14}GeV\) and
\(m_{\phi_W}=9.31\times10^{13}GeV\). They are ghosts.
The masses of W and Z and propagators
of W and Z fields are determined to be
\[m^2_W={1\over2}g^2 m^2_t,\;\;\;
m^2_Z={1\over2}(g^2+g^{'2})m^2_t,\;\;\;
G_F=\frac{1}{2\sqrt{2}m^2_t}.\]
where $m_t$ is the top quark mass. They agree with data very well.
\[\Delta^Z_{\mu\nu}=
\frac{1}{p^2-m^2_Z}\{-g_{\mu\nu}+(1+\frac{1}{2\xi_Z})\frac{p_\mu p_\nu}{
p^2-m^2_{\phi_Z}}\},\]
\[\Delta^W_{\mu\nu}=
\frac{1}{p^2-m^2_W}\{-g_{\mu\nu}+(1+\frac{1}{2\xi_W})\frac{p_\mu p_\nu}{
p^2-m^2_{\phi_W}}\},\]
\(\xi_Z=-1.18\times10^{-25}\) and
\(\xi_W=-3.73\times10^{-25}\).
The dynamically generated scalars behave like Higgs. In this new EW theory the Faddeev-Popov
procedures are not required.


\begin{thebibliography}{0}
\bibitem{} B.A.Li, Phys. Rev. {\bf D52}, 5165(1995); 5184(1995); J.Gao and B.A.Li, Phys.Rev.
{\bf D61},113006(2000); B.A.Li and J.X.Wang, Phys.Lett.{\bf B543},48(2002);
B.A.Li,D.N.Gao, and M.L.Yan,Phys.Rev.{\bf D58}094031(1998);
B.A.Li,Phys.Rev.{\bf D55},1425(1997); 1436(1997); EPJ,{\bf A10},347(2001).
\bibitem{} B.A.Li, Proceedings of ICHEP2004, Beijing, China 16-22 Aug., 2004, p.764.
edited by H.Chen, D.S.Du,W.Li, and C.D.Lu.
\bibitem{} B.A.Li and Y.L.Wu, hep-ph/0509054.
\bibitem{} Bing An Li and Tu Nan Yuan, Atomic Energy, {\bf 7-8}460(1966).
\bibitem{} B.A.Li, Intern. J of Modern Physics {\bf A19},4813(2004); {\bf A25},
3607(2002);
{\bf A17},2417(2002); {\bf A16},4171(2001).
\end{thebibliography}
\end{document}